\documentclass[preprint,prd,aps,tighten,nofootinbib,amssymb]{revtex4}
\usepackage{color}
\input{colordvi.tex}
\usepackage{amsmath}
\usepackage[dvips]{graphicx}
\usepackage{subfigure}
\newcommand{\beq}{\begin{equation}}
\newcommand{\eeq}{\end{equation}}
\newcommand{\beqa}{\begin{eqnarray}}
\newcommand{\eeqa}{\end{eqnarray}}

\newcommand{\CO}{{\cal O}}
\newcommand{\vecs}[1]{\mbox{\boldmath${#1}$}}

\begin{document}
\widetext
\draft

\begin{flushright}
DESY 11-182 \\
UT-11-34\\
RESCEU-31/11
\end{flushright}

\title{Phase transition and monopole production in \\  supergravity inflation }

\author{Kohei Kamada}%
\affiliation{Deutsches Elektronen-Synchrotron DESY, Notkestra\ss e 85, D-22607 Hamburg, Germany}

\author{Kazunori Nakayama}
\affiliation{ Department of Physics, Graduate School of Science, 
The University of Tokyo, Tokyo 113-0033, Japan}

\author{Jun'ichi Yokoyama}
\affiliation{Research Center for the Early Universe (RESCEU),
Graduate School of Science, The University of Tokyo, Tokyo 113-0033, Japan}
\affiliation{Institute for the Physics and Mathematics of the Universe (IPMU),
The University of Tokyo, Kashiwa, Chiba, 277-8568, Japan}

\date{\today}

\begin{abstract}
In F-term supergravity inflation models, scalar fields other than
the inflaton generically receive a Hubble-induced mass, which may
restore gauge symmetries during inflation and phase transitions
may occur during or after inflation as the Hubble parameter decreases.
We study monopole (and domain wall) production associated with
such a phase transition in chaotic inflation in supergravity and
obtain a severe constraint on the symmetry breaking scale which
is related with the tensor-to-scalar ratio.
Depending on model parameters, it is possible that monopoles are
sufficiently diluted to be free from current constraints but still
observable by planned experiments.
\end{abstract}

\maketitle

\section{Introduction}

Although the standard model (SM) of particle physics has been tested with great accuracy, 
there remain many issues that the SM cannot explain, such as the origin of dark matter or the hierarchy 
between the electroweak scale and the Planck scale. Thus, we need physics beyond the SM. 
One of the most promising candidates is supersymmetry (SUSY)~\cite{Nilles:1983ge}, which can solve the above problems 
naturally. Moreover, the running of gauge coupling constants in SUSY suggests the unification of gauge interactions. 

In the grand unified theories (GUTs)~\cite{Langacker:1980js,Slansky:1981yr}, the gauge interaction is 
described by a gauge group $G_{\rm GUT}$ with a single gauge coupling constant, which contains the standard model 
gauge group, $G_{\rm SM}=SU(3)_C \times SU(2)_L \times U(1)_Y$, as a subgroup.  
Thus far, many models of GUTs, especially its supersymmetric version (SUSY GUTs) have been proposed, 
such as those based on $SU(5)$ or $SO(10)$. 
Some of them directly break down to $G_{\rm SM}$ and others have one or more intermediate 
symmetry groups between $G_{\rm GUT}$ and $G_{\rm SM}$. 
Phenomenological aspects of GUTs have been studied intensively~\cite{Langacker:1980js}. 
The idea of GUT has also opened a window for studying the early Universe such as baryogenesis~\cite{yoshimura1978}.  

However, there is a severe problem in GUTs. 
When the GUT gauge symmetry and other intermediate symmetry 
breaks down, topological defects~\cite{stringreveiw} such as (magnetic) monopoles 
\cite{'tHooft:1974qc,Polyakov:1974ek}, strings or domain walls 
are formed through the Kibble mechanism~\cite{Kibble:1976sj}. 
In particular, magnetic monopoles are inevitably produced during the course of GUT phase transition 
down to $U(1)_Y$. 
They are copiously produced and overclose the Universe\footnote{This problem is also discussed 
in the context of hybrid inflation models based on SUSY GUTs~\cite{hybrid,su5inf,antusch}.}, 
although the detail of monopole production depends on the pattern of symmetry breaking~\cite{Jeannerot:2003qv}. 

Inflation~\cite{Sato:1980yn,Linde:2005ht} was proposed as a solution to this problem because 
inflationary expansion of the Universe dilutes monopoles sufficiently. 
It can also solve other cosmological problems such as horizon and flatness problems and account for the origin of 
primordial fluctuations. Now it is a part of ``standard'' cosmology. 
In order for the monopoles to be diluted sufficiently, inflation must take place after the phase transition.
On the other hand, scenarios in which phase transition takes place after inflation are ruled out unless 
one assumes an artificial mechanism to dilute them such as thermal inflation~\cite{Yamamoto:1985rd,Lazarides:2000em}. 
In order for the GUT symmetry not to be restored after inflation,
the reheating temperature after inflation must be much smaller than the GUT scale.
Otherwise, thermal effects might stabilize the Higgs at the origin where the symmetry is restored.

In this paper, we emphasize that not only the reheating temperature, but also the Hubble scale during inflation
must be smaller than the GUT scale in order to avoid the monopole problem
in the context of F-term inflation models in supergravity~\cite{Lyth:1998xn}.
A typical feature of F-term inflation models is the appearance of the 
Hubble-induced mass~\cite{Dine:1995uk} to any scalar field, which is 
inevitable unless one assumes some shift symmetries~\cite{Goncharov:1983mw,Kawasaki:2000yn} 
or non-canonical K${\rm {\ddot a}}$hler  potential.  
At the onset of inflation, the Hubble parameter is very large and gradually decreases during inflation. 
The Higgs field responsible for the symmetry breaking also acquires the Hubble-induced mass, 
and it may be this Hubble mass term that controls the phase transition~\cite{Freese:1995vp} 
in the way quite similar to the curvature induced phase transitions proposed 
in Refs.~\cite{Yokoyama:1988zza,Yokoyama:1989xj,Nagasawa:1991zr}. 
In particular, it is possible that the phase transition takes place during inflation\footnote{Scenarios 
where phase transition takes place at a late stage of inflation 
by introducing a coupling between the Higgs field and inflaton~\cite{Lazarides:1984pq,Shafi:1984tt} 
have also been proposed,
although some of which focused on the cosmic string formation. }.
This consideration leads us to the conclusion that the symmetry breaking scale
must be larger than the Hubble scale of inflation.

Among many (F-term) inflation models, chaotic inflation 
\cite{Linde:1983gd,Kawasaki:2000yn} is one of the most interesting models from this viewpoint.
It is one of the simplest models of inflation and predicts large tensor perturbation 
that can be detected by PLANCK~\cite{PLANCK_HP,PLANCK_coll}, with its 
energy scale close to the GUT scale $\sim 10^{16} {\rm GeV}$. 
Therefore, if future detection of B-mode polarization confirms the chaotic inflation,
the symmetry breaking scale will be tightly constrained.
Moreover, it may be expected that the spontaneous symmetry breaking of GUT, 
or other intermediate symmetries, takes place {\it during} chaotic inflation.
In such a case, the expected monopole flux may be within the reach of future experiments.

In this paper, we focus on the chaotic inflation model 
\cite{Kawasaki:2000yn,Linde:1983gd} and study the feature of phase transition in detail.  
We find that the symmetry breaking scale associated with monopole production must be 
large enough so that it occurs well before the end of inflation. 
We carefully estimate the number density of monopoles at the time of the phase transition during chaotic inflation. 
We find that the symmetry breaking scale $M$ is bounded below as $ M > (1-4) \times 10^{13} \kappa^{-1}{\rm GeV}$
depending on the reheating temperature, with $\kappa$ being the coupling constant in the model,
if the $B$-mode of the CMB polarization is detected by on-going and future experiments such as 
PLANCK \cite{PLANCK_HP,PLANCK_coll}, QUIET+PolarBeaR \cite{Hazumi:2008zz}
or LiteBIRD~\cite{Hazumi:2008zz,LiteBird}.
If the reheating temperature of the Universe is determined by precise measurements of CMB~\cite{Martin:2010kz} 
or the detection of inflationary gravitational waves~\cite{Nakayama:2008ip}, this constraint becomes severer. 
Moreover, we find that there are parameter space where the direct detection of monopoles can be expected. 

This paper is organized as follows. 
In the next section, we make a general discussion on the phase transition during inflation. 
In Sec.~\ref{sec:4}, we investigate the case of chaotic inflation and monopole production quantitatively. 
The final section is devoted to conclusions and discussion.

\section{Phase transition in F-term inflation \label{sec:3}}

Monopoles are topological defects that are formed when a gauge group $G$ breaks down to 
a subgroup $H$ of $G$ if the homotopy group satisfies $\pi_2(G/H) \not = 0$. 
The spontaneous breaking of GUT and other intermediate symmetries down to the SM gauge group
generally predict monopole production~\cite{Jeannerot:2003qv}. 

When a scalar multiplet that has a gauge charge of $G$ acquires a  
nonvanishing expectation value, spontaneous symmetry breaking takes place.  
To be concrete, let us consider following superpotentials for the Higgs field,
\begin{equation}
W=\kappa S \left({\rm Tr} \Sigma^2 - M^2 \right),  \label{sp1}
\end{equation}
where the supermultiplet $\Sigma$ is adjoint representation of $G=SU(N)$, or, 
\begin{equation} 
W=\kappa S \left(H {\bar H} - M^2 \right),  \label{sp2}
\end{equation}
where the supermultiplets $H ({\bar H})$ are  (anti-) fundamental representations of $G$, or 
\begin{equation} 
W=\kappa S \left(\Phi {\bar \Phi} - M^2 \right),  \label{sp3}
\end{equation}
where supermultiplets $\Phi ({\bar \Phi})$ are (anti-)fundamental representations of $G^\prime$ 
with $G=G^\prime \times U(1)$ and oppositely charged under $U(1)$ and so on. 
Here $S$ is an additional singlet, $\kappa$ is a numerical coefficient, which is taken to be real and positive,
less than $\CO(1)$, and $M$ is a mass parameter corresponding to the symmetry breaking scale. 
If the system has a monopole solution, there are at least three scalar degrees of freedom 
whose potential can be reduced to 
\begin{equation}
V=\frac{1}{2}\kappa^2 \left(\sum_{a=1}^3 \sigma_a^2 -2M^2\right)^2, \label{scpot}
\end{equation}
after field redefinition and imposing D-flat condition. 
Here $\sigma_a (a=1,2,3)$ is real scalar degrees of freedom. 
Equation \eqref{scpot} has a vacuum at $S=0, \sum_a \sigma_a^2=2M^2$, around which 
the $G$ is spontaneously broken. 
An illustrative example is a left-right symmetric group $SU(2)_L \times SU(2)_R \times U(1)_{B-L}$,
broken by the VEV of an adjoint representation $\Sigma$ of $SU(2)_R$,
which is further followed by the successive phase transition
$U(1)_R \times U(1)_{B-L} \rightarrow U(1)_Y$.
This discussion is general in all the symmetry breaking that predicts monopole production. 
More concrete models will be discussed in Sec. \ref{sec:5}.  
Hereafter we mainly use $\Sigma$ as a Higgs multiplet symbolically but the result is general. 

Before going into the discussion of phase transition during inflation, let us briefly see the phase transition 
triggered by thermal effects.
The usual thermal phase transition proceeds as follows.
At a high temperature, the Higgs field acquires a large thermal mass of order of $\sim T^2 \sigma_a^2$ 
and hence it is stabilized at the origin. 
As the temperature of the Universe decreases, the bare tachyonic mass $-\kappa^2 M^2\sigma_a^2$ overwhelms 
thermal mass. 
At that time the Higgs field becomes unstable and phase transition takes place producing monopoles. 
The monopole mass is given by \cite{'tHooft:1974qc,Polyakov:1974ek}, 
\begin{equation}
M_{\rm m} \simeq \frac{4 \pi M}{g_{\rm G}},   \label{massmp}
\end{equation}
where $g_{\rm G}$ is the gauge coupling constant. 
The monopole is so heavy that we can neglect the effect of monopole annihilation~\cite{Yokoyama:1988zza}.
Without a dilution mechanism~\cite{Langacker:1980kd}, 
they would soon overclose the Universe.
In order to avoid the monopole overproduction in thermal phase transition,
the reheating temperature after inflation cannot be higher than the GUT
scale.
In SUSY, however, 
there is an even more strict upper bound on the reheating temperature as $T_R \lesssim 10^{6-9}$\,GeV
for avoiding the overproduction of the gravitino~\cite{Kawasaki:2008qe}.
Thus the GUT symmetry is likely never restored thermally after inflation.

Instead,  the symmetry may be restored and the 
phase transition may be triggered by the Hubble-induced
mass, which we focus on hereafter. 
Assuming canonical K${\rm {\ddot a}}$hler potential for the Higgs multiplet(s), the scalar potential of the 
system includes a Hubble-induced mass 
\cite{Dine:1995uk}, 
\begin{equation}
V(\sigma_a) \ni \exp\left(\frac{|\Sigma|^2}{M_G^2} \right) \left|\frac{\partial W_{\rm inf}}{\partial \phi_{\rm inf}}\right|^2 \simeq \frac{3}{2} H^2 \sigma_a^2, 
\end{equation}
where $\phi_{\rm inf}$ is the inflaton, $W_{\rm inf}$ is its superpotential and $M_G$ is the reduced Planck mass\footnote{
	Here $\phi_{\rm inf}$ should be regarded as the field whose F-term dominates the potential energy during inflation.
	It necessarily does not coincide with the inflaton in a usual sense, which is a slowly rolling scalar field in the potential.
}. 
Here we have used the Friedmann equation
\begin{equation}
3 H^2 M_G^2 = V(\phi_{\rm inf})=\left|\frac{\partial W_{\rm inf}}{\partial \phi_{\rm inf}}\right|^2. 
\end{equation}
If $\kappa M < H$ at the end of inflation, the symmetry is not broken until the Hubble parameter 
decreases to $\sim \kappa M$.
Thus, monopoles are produced after inflation, which leads to a cosmological disaster. 
On the other hand, if $\kappa M \gg H$ at the end of inflation, the symmetry is broken before the end of inflation 
and monopoles can be diluted sufficiently. 
Moreover, it is possible that the phase transition takes place just before the end of inflation
if the Hubble parameter decreases with a sizable rate during inflation.

This argument offers us an important suggestion. 
In supersymmetric F-term inflation models, the symmetry breaking scales of 
GUT and other intermediate symmetries
that are associated with monopole (or domain wall) production
{\it must} be larger than the Hubble parameter during inflation. Note that GUT gauge group $G_{\rm GUT}$ 
does not have to break directly down to $G_{\rm SM}$ but it is possible that there are one or more 
gauge groups between $G_{\rm GUT}$ and $G_{\rm SM}$, as has been mentioned.  
Therefore, there can be several breaking scales below the GUT scale ($\sim 10^{16}$ GeV). 
All these symmetry breaking scales are constrained by this condition,
if the corresponding symmetry breaking is associated with monopole or domain wall production.

Actually, the Hubble parameter during inflation can be proved by the observation of $B$-mode polarization in CMB. 
Tensor perturbation in the primordial perturbation, which is imprinted in the B-mode polarization,
is related to the Hubble parameter during inflation as 
\begin{equation}
{\cal P}_T=\frac{8}{M_G^2}\left. \left(\frac{H}{2\pi}\right)^2 \right|_{H=k/a}. 
\end{equation}
If the satellite experiments such as Planck \cite{PLANCK_HP,PLANCK_coll} 
or LiteBird \cite{Hazumi:2008zz,LiteBird}, or the ground-based detectors 
such QUIET+PolarBeaR \cite{Hazumi:2008zz} detect the $B$-mode of the CMB polarization in the near future, 
GUT and other intermediate breaking scales will be severely constrained as
$M > 10^{13} \kappa^{-1}$ GeV $( 10^{-12} \kappa^{-1} {\rm GeV})$ for $r=0.1 (10^{-3})$ . 

Interestingly enough, 
as noted earlier, the phase transition may take place just before the end of inflation.
In such a case the dilution of monopoles is rather mild, and an observable amount of monopoles may be left in the Universe.
The precise constraint on the symmetry breaking scale $M$ depends on 
when the phase transition took place.
It is interesting if the phase transition takes place slightly before the end of inflation,
because monopole searches such as IceCube~\cite{Hardtke:2007zz}, 
combined with the detection of B-mode, will provide us with useful information on the GUT symmetry breaking.
We will investigate in detail the possibility of the phase transition and monopole production during inflation  
in the next section.

Note that there is much literature that discusses the monopole problem associated with 
the superpotential of the form Eqs.~\eqref{sp1}, \eqref{sp2}, or \eqref{sp3}, especially in the 
inflation models embedded in the Higgs sector of the GUT symmetry breaking. For example, 
in Ref.~\cite{su5inf}, the shifted inflation is embedded in the SU(5) GUT model and 
the monopole problem is avoided by violating the SU(5) gauge 
symmetry during the course of inflation already. 
In Ref. \cite{antusch}, a flat direction in GUT (to be concrete, they choose Pati-Salam and SO(10) models) 
is identified with the inflaton  for hybrid inflation and GUT symmetry is broken already during inflation thanks to 
the higher  dimensional operators. 
In Ref. \cite{Pallis:2011gr}, nonminimal ``Higgs'' inflation is embedded in the Pati-Salam model and 
the monopole problem is avoided by identifying the Higgs field with the inflaton for the chaotic inflation, 
which means that Pati-Salam gauge symmetry is already broken during the course of inflation. 
Other recent studies are listed in Ref.~\cite{infmonorefs}. 
Generally, however, inflation does not need to be embedded in the GUT-breaking 
sector\footnote{Inflation models embedded in the SUSY-breaking sector are discussed in, for example, 
Ref.~\cite{Craig:2008tv}. }. In this case, GUT symmetry restoration during inflation due to the large Hubble-induced 
mass and succeeding phase transition is inevitable if the Hubble parameter during inflation is large enough 
as we have seen above. 
Moreover, since the symmetry preserving state must not be the local potential minimum, superpotential 
of the form Eqs.~\eqref{sp1}, \eqref{sp2}, or \eqref{sp3} is needed. 
Here, we focus on such a case although the superpotential has a similar form to that for the hybrid inflation.

\section{Monopole production in chaotic inflation \label{sec:4}}

\subsection{Monopole Production during Chaotic Inflation}

In this section, we investigate monopole production during chaotic inflation in supergravity~\cite{Kawasaki:2000yn}.
The energy scale at the end of chaotic inflation is around $10^{13}$ GeV, which is close to the reasonable GUT and 
other intermediate symmetry breaking scales. 
Therefore, it is worth focusing on this specific inflation model. 

The K${\rm {\ddot a}}$hler and superpotential for the model we adopt here are 
\begin{align}
K&=\frac{1}{2}(\Phi+\Phi^\dagger)^2+|X|^2+|S|^2+|\Sigma|^2, \label{Kahler} \\ 
W&=m X \Phi + \kappa S \left({\rm Tr} \Sigma^2 -M^2\right),    \label{super}
\end{align}
where $\Phi$ is the inflaton and $X$ is an additional singlet,
$m$ is the inflaton mass and $M$ is the symmetry breaking scale.
We impose $R$-symmetry and discrete $Z_2$ symmetry
in order to suppress all other unwanted couplings such as $X\Phi^2, S\Phi,$ etc\footnote{
Notice that the term $\epsilon S \Phi^2$ in the superpotential is allowed by these symmetries 
but it breaks the shift symmetry of $\Phi$.
Taking the inflaton mass $m$ as an order parameter of the shift symmetry breaking,
we expect that the coupling constant $\epsilon$ is suppressed enough, say $\epsilon\sim m^2\sim 10^{-10}$
in Planck units. Thus, the following discussion does not change. }. 
Charge assignments on the fields are shown in Table~I.
In this model, the imaginary part of $\Phi$, $\varphi\equiv{\rm Im} \Phi /\sqrt{2}$ acts as the inflaton 
because the shift symmetry in the K${\rm {\ddot a}}$hler potential, $\Phi \rightarrow \Phi+i c$ where $c$ 
is a real parameter, protects it from obtaining the exponential growth of the scalar potential.
Explicitly, the scalar potential includes a term like
\begin{equation}
V=e^K\left[D_i W K^{i {\bar j}} D_{\bar j} W^* -\frac{3}{M_G^2}|W|^2\right] \ni 
3 H^2 M_G^2 \exp\left( \frac{2({\rm Re} \Phi)^2}{M_G^2}\right), 
\end{equation}
where 
\begin{equation}
D_i W \equiv \frac{\partial W}{\partial \phi_i}+\frac{1}{M_G^2} \frac{\partial K}{\partial \phi_i} W,  \quad K^{i{\bar j}} \equiv \left(\frac{\partial ^2 K}{\partial \phi_i \partial \phi_j^*}\right)^{-1}. 
\end{equation}
This potential has a vacuum at 
\begin{equation}
\Phi=X=S=0, \quad \sum_{a=1}^3 \sigma_a^2=2M^2,   \label{vacuum}
\end{equation}
where $\sigma_a$ is the effective real scalar degree of freedom of the Higgs field $\Sigma$. 
Around this minimum, the Higgs fields have a mass of $\kappa M$ 
and massive gauge bosons acquire a mass of $g_{\rm G} M$.

\begin{table}[t!]
  \begin{center}
    \begin{tabular}{ | c | c | c | c | c |}
      \hline 
         ~          &  $\Phi$  & $X$ & $S$ & $\Sigma$ \\ \hline
        $R$ & 0 & $+2$ & $+2$ & 0 \\ \hline
        $Z_2$  & $-1$ & $-1$ & 0 & 0 \\ \hline   
    \end{tabular}
    \caption{ 		
     	Charge assignments on superfields in the model under 
	the $R$-symmetry and $Z_2$-symmetry.
     }
  \end{center}
  \label{table:charge}
\end{table}

The inflaton $\varphi$ has a simple quadratic potential beyond the Planck scale.
If $\varphi$ acquires a large field value, $\varphi \gg M_G$, 
it enters the slow-roll regime and inflation takes place.
The equation of motion is given by
\begin{align}
&3H (\varphi) {\dot \varphi}+m^2 \varphi=0,  \\
&3H^2 (\varphi)M_G^2 = \frac{1}{2}m^2 \varphi^2, 
\end{align}
The slow-roll condition is violated at $\varphi \simeq \varphi_{\rm e} \equiv \sqrt{2} M_G$, 
when inflation is terminated. 
The number of $e$-folds of inflation from $\varphi$ to $\varphi_{\rm e}$ reads 
\begin{equation}
{\cal N}(\varphi)=\frac{1}{M_G^2} \int_{\varphi_{\rm e}}^\varphi \frac{V}{V^\prime} d\varphi =\frac{1}{4M_G^2}(\varphi^2-2 M_G^2). 
\end{equation}
Observable quantities, such as the magnitude of the power spectrum of the curvature perturbation ${\cal P}_{\cal R}$, 
the scalar spectral index of primordial curvature perturbation, $n_s$,  and 
the tensor-to-scalar ratio, $r$, are expressed in terms of the number of $e$-folds when observable scales exit the horizon
as follows, 
\begin{align}
{\cal P}_{\cal R} &= \frac{1}{24 \pi^2 \epsilon}\frac{V}{ M_G^4} \simeq \frac{4m^2}{3 \pi^2 M_G^2}  {\cal N}_{\rm COBE}^4, \\
n_s-1&=-6 \epsilon+2 \eta\simeq -\frac{2}{{\cal N}_{\rm COBE}}, \\
r&=16 \epsilon\simeq  \frac{8}{{\cal N}_{\rm COBE}}. 
\end{align}
where ${\cal N}_{\rm COBE}\simeq 50-60$ depending on the reheating temperature.
Here $\epsilon=(M_G^2/2) (V^\prime/V)^2$ and $\eta=M_G^2 (V^{\prime \prime}/V)$ are slow-roll parameters. 
The present observation, ${\cal P}_{\cal R} = 2.4 \times 10^{-9}$ 
 \cite{Komatsu:2010fb}, determines the mass of the inflaton to be $m\sim 10^{13}$ GeV. 
The tensor-to-scalar ratio is predicted to be $r\sim 0.13-0.16$, which is expected to be detected by 
PLANCK~\cite{PLANCK_HP,PLANCK_coll}.  

During inflation, the relevant part of the scalar potential is given by\footnote{
	If there exists a nonminimal K\"ahler potential like $k |X|^2 |\Sigma|^2 / M_G^2$,
	the Higgs field receives an additional Hubble mass correction.
	This does not modify the following arguments as long as $|k| \lesssim 1$.
}
\begin{equation}
V=\frac{1}{2}m^2 \varphi^2+ m^2|X|^2+3H^2 ( \varphi)\left( \frac{1}{2}\eta^2 +|S|^2+\frac{1}{2} \sum_a \sigma_a^2 \right) + \frac{1}{2}\kappa^2 \left( \sum_a \sigma_a^2-2M^2 \right)^2. 
\end{equation}
The mass eigenvalue of the Higgs fields around the origin reads
\begin{equation}
m_{\sigma }^2=3H^2 (\varphi) - \kappa^2 M^2. \label{higgsmass}
\end{equation}
If the Hubble parameter is large enough, all the fields except for $\varphi$ and $X$ 
quickly settle down to the origin because of the Hubble-induced masses. 
Then the gauge symmetry is restored. 
$X$ also has a mass of $m$ and hence its evolution is described by the slow-roll equation. 
However, in the case where $\varphi \gg |X|$, which we consider here, it does not
influence the inflaton dynamics and the density perturbations~\cite{Kawasaki:2000yn}.  

Now we see how the phase transition and monopole production proceed,
following the arguments of Ref.~\cite{Nagasawa:1991zr}. 
When the Hubble parameter is large enough, the Higgs fields settle down to the origin. 
As the Hubble parameter decreases, the mass eigenvalue of Higgs fields [Eq. \eqref{higgsmass}] becomes negative. 
As the symmetric state $\Sigma=0$ becomes unstable, it starts to fall down towards 
the true vacuum. 
One may expect that the phase transition takes place when 
the minus of the Higgs mass squared becomes as large as the Hubble parameter, 
\begin{equation}
m_{\sigma }^2 = V^{\prime \prime}(\Sigma=0)= -H^2(\varphi), 
\end{equation}
when the slow-roll conditions for the Higgs fields are violated and their dynamics is governed by the classical potential force.  
However, in the present case where the effective mass is time dependent, 
this treatment may not be valid. 
In order to treat the behavior of the Higgs fields around the epoch of phase transition appropriately, 
we adopt the stochastic approach~\cite{Stoch,Starobinsky:1994bd}. 
When the Higgs field is in the slow-roll regime with $|V^{\prime \prime}(\Sigma = 0)|\lesssim H^2(\varphi)$, 
a coarse-grained or a long-wavelength mode of the Higgs field 
obeys the Langevin equation~\cite{Stoch,Starobinsky:1994bd}, 
\begin{equation}
\frac{d \sigma ({\vecs x},{\cal N})}{d{\cal N}}=-\frac{V^\prime(\sigma)}{3 H^2({\cal N})}+\frac{f({\vecs x},{\cal N})}{H({\cal N})},  \label{Lange1}
\end{equation}
where ${\cal N}\equiv \log a(t)-\log a(t_0)$ is the number of $e$-folds from $t_0$ to $t$ 
($t_0$ is an initial time that can be taken arbitrarily) 
and $f({\vecs x},{\cal N})$ is a stochastic noise whose correlation function is given by 
\begin{equation}
\langle f({\vecs x},{\cal N}_1) f({\vecs x},{\cal N}_2) \rangle = \frac{H^4({\cal N}_1)}{4 \pi^2}\delta ({\cal N}_1-{\cal N}_2), \quad  \langle f({\vecs x},{\cal N})\rangle =0. 
\end{equation}
The first term in Eq. \eqref{Lange1} represents the classical force and the second one represents the stochastic force. 
When the first term overwhelms the second term, 
\begin{equation}
\left|\frac{V^{\prime}(\langle \sigma^2({\cal N}^\prime)\rangle ^{1/2})}{3 H^2 ({\cal N})}\right| \gg \frac{\langle f^2({\cal N})\rangle^{1/2} }{H({\cal N})} = \frac{H({\cal N})}{2 \pi }, \label{class}
\end{equation}
the classical equation of motion begins to determine the dynamics of the Higgs field. 
After that, its dynamics is decisive. 
We expect that monopole distribution is determined at this time. 

Noting that the Hubble parameter is written by 
\begin{equation}
H^2({\cal N})=H_0^2-\frac{2{\cal N}}{3}m^2, 
\end{equation}
where $H_0^2=m^2 \varphi(t_0)^2/2$, Eq.~\eqref{Lange1} reads 
\begin{equation}
\frac{d \sigma ({\vecs x},{\cal N})}{d{\cal N}}=-\left(1-\frac{\kappa^2 M^2}{3 H_0^2 - 2 {\cal N} m^2}\right) \sigma  ({\vecs x},{\cal N})+\frac{f({\vecs x},{\cal N})}{\sqrt{H_0^2-\dfrac{2{\cal N}}{3}m^2}}. \label{Lange2}
\end{equation}
Equation \eqref{Lange2} is solved as 
\begin{align}
\sigma({\cal N})=&\left\{\sigma({\cal N}= 0)+\int_0^{\cal N} d{\cal N^\prime}\frac{f({\vecs x},{\cal N}^\prime)}{\sqrt{H_0^2-\dfrac{2{\cal N^\prime}}{3}m^2}}\exp\left[\int_0^{{\cal N}^\prime} d{\cal N}^{\prime\prime}\left(1-\frac{\kappa^2 M^2}{3 H_0^2 - 2 {\cal N^{\prime\prime}} m^2}\right) \right]\right\} \notag \\
& \times \exp\left[-\int_0^{\cal N} d{\cal N^\prime} \left(1-\frac{\kappa^2 M^2}{3 H_0^2 - 2 {\cal N^\prime} m^2}\right) \right] \notag \\
=& \left[\sigma_{0}+\int_0^{\cal N} d{\cal N^\prime}\frac{f({\vecs x},{\cal N})}{H_0} \left(1-\frac{2 m^2 {\cal N}^\prime}{3H_0^2}\right)^{\kappa^2 M^2/2 m^2-1/2} e^{{\cal N}^\prime}\right] \notag \\
&\times \left(1-\frac{2m^2 {\cal N}}{3 H_0^2}\right)^{-\kappa^2 M^2/2m^2} e^{-{\cal N}}. 
\end{align}
From this solution, we can follow the evolution of the expectation value of the Higgs field, 
\begin{align}
\langle \sigma^2 ({\cal N})\rangle &= \left[\langle \sigma_{0}^2 \rangle +\frac{H_0^2}{4 \pi^2}\int_0^{\cal N} d{{\cal N}^\prime} \left(1-\frac{2 m^2 {\cal N}^\prime}{3H_0^2}\right)^{\kappa^2 M^2/ m^2+1} e^{2{\cal N}^\prime}\right]\left(1-\frac{2m^2 {\cal N}}{3 H_0^2}\right)^{-\kappa^2 M^2/m^2} e^{-2{\cal N}}. 
\end{align}
Expanding the following terms as 
\begin{align}
\log \left[ e^{-2{\cal N}} \left(1-\frac{2 m^2 {\cal N}}{3 H_0^2}\right)^{-\frac{\kappa^2 M^2}{m^2}}\right] =& \left(\frac{2 \kappa^2 M^2}{3 H_0^2}-2\right){\cal N} +\frac{2 \kappa^2 m^2 M^2}{9 H_0^4}{\cal N}^2\notag \\
&+\frac{\kappa^2 M^2}{m^2}{\cal O}\left(\left(\frac{2m^2 {\cal N}}{3H_0^2}\right)^3\right), \\
\log \left[ e^{2{\cal N}} \left(1-\frac{2 m^2 {\cal N}}{3 H_0^2}\right)^{\frac{\kappa^2 M^2}{m^2}+1}\right] =&2\left(1-\frac{\kappa^2 M^2+m^2}{3 H_0^2}\right){\cal N} -\frac{2 m^2}{9 H_0^4}(m^2+\kappa^2 M^2){\cal N}^2  \notag \\
&+\left(\frac{\kappa^2 M^2}{m^2}+1\right){\cal O}\left(\left(\frac{2m^2 {\cal N}}{3H_0^2}\right)^3\right), 
\end{align}
and setting 
\begin{equation}
H_0^2=\frac{\kappa^2 M^2}{3}, 
\end{equation}
corresponding to $m^2_\sigma(H_0^2)=0$, 
one obtains a following approximation, 
\begin{align}
\langle \sigma^2 ({\cal N})\rangle \simeq &\left\{ \langle \sigma_{0}^2 \rangle + \frac{\kappa^2 M^2}{8\sqrt{2}\pi^{3/2} } \frac{ce^{1/2(1+c)}}{\sqrt{2 (1+c)}} \left[{\rm erf}\left(\frac{\sqrt{2(1+c)}}{c} {\cal N}+\frac{1}{\sqrt{2(1+c)}}\right)- {\rm erf}\left(\frac{1}{\sqrt{2(1+c)}}\right)\right] \right\} \notag \\
&\times \exp \left(\frac{2}{c} {\cal N}^2\right), 
\end{align}
where we have defined $c \equiv \kappa^2 M^2/m^2$. 
Using the approximate expression, 
\begin{equation}
{\rm erf} (x)\simeq 
\begin{cases}
x \quad (x \ll 1) , \\
1 \quad (x \gg 1)
\end{cases}
\end{equation}
we find
\begin{equation}
\langle \sigma^2 ({\cal N})\rangle \simeq \left( \langle \sigma_{0}^2 \rangle + \frac{m^2}{8 \sqrt{2} \pi^{3/2} } \frac{ c^3 e^
{1/2(1+c)}}{\sqrt{2 (1+c)}} \right)\exp \left(\frac{2 }{c} {\cal N}^2\right), 
\end{equation}
for ${\cal N} \gg  \sqrt{c/2}$. 
Thus $\langle \sigma^2 ({\cal N})\rangle$ starts to grow exponentially 
at that time and soon satisfies the condition Eq.~\eqref{class}. 
Therefore, we conclude that the dynamics of the Higgs field enters the classical regime 
at ${\cal N} \simeq \sqrt{c/2}$. 
\footnote{Here we have neglected the quartic term in the potential. This can be validated when 
$\kappa \ll 8\sqrt{2} \pi^{3/2} c^{3/2}$. This condition is derived from the condition that the minimum of the 
Higgs potential at the number of $e$-folds ${\cal N}$ is larger than 
$\langle \sigma^2({\cal N})\rangle^{1/2}$ at small ${\cal N}$. }
This corresponds to the Hubble parameter 
\begin{equation}
H_{\rm f}\equiv H(t_{\rm f})\simeq \frac{m}{\sqrt{3}}\left(c-\sqrt{2c}\right)^{1/2}, 
\end{equation}
where $t_{\rm f}$ is defined as the monopole formation time. 
Inflation continues after the phase transition. 
The number of $e$-folds thereafter reads 
\begin{equation}
\Delta {\cal N} =\frac{3H_{\rm f}^2}{2m^2}-\frac{1}{2} \simeq \frac{1}{2}(c-\sqrt{2c}-1).  \label{efolaft}
\end{equation}

Next we estimate the power spectrum of the distribution of the Higgs field in order to 
estimate the number density of monopoles. 
Naively, one may assume that the mean separation of monopoles would be the Hubble length at its 
formation time. 
However, due to the inflationary expansion, scalar fields are correlated beyond the horizon scale, 
and hence its mean separation becomes larger than the Hubble length. 
The mode function of the Higgs field obeys the equation of motion, 
\begin{equation}
{\ddot \sigma}_{k}(t)+3 H {\dot \sigma}_{k}(t)+\left(\frac{k}{a(t)}\right)^2 \sigma_{k}(t)+m_{\sigma}^2 (t) \sigma_{k}(t)=0. \label{modeeq}
\end{equation}
Defining a variable as ${\tilde \sigma}_{k} \equiv a^{3/2}(t) \sigma_{k}$, Eq. \eqref{modeeq} can be rewritten as 
\begin{equation}
{\ddot {\tilde \sigma}_{k}} + \left[\left(\frac{k}{a(t)}\right)^2+m_{\sigma}^2 (t) -\frac{3}{2}{\dot H}-\frac{9}{4} H^2\right]{\tilde \sigma}_{k}=0. \label{modmodeeq}
\end{equation}
In the inflationary stage, the condition $|{\dot H}| \ll H^2$ is satisfied and hence we can 
approximate the Hubble parameter $H$ as a constant over several expansion time scales.
Then, we have 
\begin{equation}
{\tilde \sigma}_{k}\simeq \sqrt{\frac{\pi}{4 H}} H_{3/2}^{(1)}\left(\frac{k}{Ha(t)}\right)
\end{equation}
for short-wave mode, $k \gg Ha(t), m_{\sigma -} (t) a(t)$. 
Here $H_{3/2}^{(1)}$ is the Hankel function of the first kind with rank 3/2 
and we have taken the positive frequency mode so that it coincides with that in the Minkowski vacuum
in the short-wavelength limit.
We have normalized ${\tilde \sigma}_{k}$ as 
\begin{equation}
{\tilde \sigma}_{k}(t){\dot {\tilde \sigma}^*_{k}}(t)-{\tilde \sigma}^*_{k}(t){\dot {\tilde \sigma}_{k}}(t)=i. \label{short}
\end{equation}
The mode with a comoving wavenumber $k$ shifts from the short-wavelength regime to long-wavelength 
regime during the course of inflation. 
Thus, the expression \eqref{short} should be connected to the solution at long-wavelength regime. 
For the long-wavelength mode $k \ll Ha(t), m_{\sigma -} (t) a(t)$, 
Eq.~\eqref{modmodeeq} can be solved by means of the WKB approximation, 
\begin{align}
{\tilde \sigma}_{k}\simeq &A_k a^{3/2}(t_k) \left(\frac{S(t_k)}{S(t)} \right)^{1/2} \exp\left[ \int_{t_k}^t S(t^\prime) dt^\prime\right]  
\end{align}
where $t_k$ satisfies $k=H(t_k) a(t_k)$ and $S$ is defined as\footnote{Here we use the fact that $H=H_0-m^2t/3$. } 
\begin{equation}
S(t)=\frac{3}{2}H\left(1+\frac{2{\dot H}}{3 H^2}-\frac{4 m_{\sigma}^2}{9 H^2}\right)^{1/2}=\left( \kappa^2 M^2-\frac{m^2}{2}-\frac{3H^2}{4}\right)^{1/2}
\end{equation}
Here, $A_k$ is a numerical constant, which is determined below.  
Here, we neglect the decaying mode. 
Note that this approximation is valid when $|{\dot S}| \ll S^2$. 
In the epoch $H\simeq \kappa M$, which we are interested in, this condition is satisfied. 

Connecting these solutions, we find 
\begin{equation}
A_k \simeq \sqrt{\frac{H^2(t_k)}{2 k^3}}. 
\end{equation}
Thus we have the mode function in the long-wavelength regime, 
\begin{equation}
\sigma_{k}\simeq   \sqrt{\frac{H^2(t_k)}{2 k^3}}\left(\frac{a(t_k)}{a(t)}\right)^{3/2} \left(\frac{S(t_k)}{S(t)}\right)^{1/2}\exp\left[\int _{t_k}^t S(t^\prime) dt^\prime \right]. 
\end{equation}
Noting that 
\begin{align}
a=a(t_k) e^{\cal N} &=a(t_k)\exp \left[ \frac{3}{2m^2} (H(t_k)^2-H(t)^2)\right], \\
\int _{t_k}^t S(t^\prime) dt^\prime =& \frac{3}{m^2}\int_{H(t)}^{H(t_k)} \left( \kappa^2 M^2-\frac{m^2}{2}-\frac{3H^{\prime 2}}{4}\right)^{1/2} dH^\prime \notag \\
= &\frac{3}{m^2} \left\{ \frac{H(t_k)}{4}\sqrt{2(2\kappa^2 M^2-m^2)-3H^2(t_k)}-\frac{H(t)}{4}\sqrt{2(2 \kappa^2 M^2-m^2)-3H^2(t)} \right. \notag \\
& -\frac{1}{\sqrt{3}}\left( \kappa^2 M^2-\frac{m^2}{2} \right) \left(\tan^{-1} \frac{H(t_k)}{\sqrt{\frac{4}{3}( \kappa^2 M^2-m^2/2)-H^2(t_k)} } \right. \notag \\
&\left. \left.  -\tan^{-1}\frac{H(t)}{\sqrt{\frac{4}{3}( \kappa^2 M^2-m^2/2)-H^2(t)}}\right)\right\} \notag \\
\simeq & \frac{3}{ m^2} \left[\sqrt{\kappa^2 M^2-\frac{m^2}{2}} (H(t_k)-H(t))-\left(\frac{H^3(t_k)-H^3(t)}{8\sqrt{\kappa^2 M^2-m^2/2}}\right)+\cdots \right], 
\end{align}
we obtain
\begin{align}
\sigma_{k}& \simeq  \sqrt{\frac{H^2(t_k)}{2 k^3}} \left(\frac{S(t_k)}{S(t)}\right)^{1/2} \exp\left[-\frac{9}{4m^2}(H^2(t_k)-H^2(t))+\frac{3 }{ m^2}\sqrt{\kappa^2 M^2 -\frac{m^2}{2}}(H(t_k)-H(t)) \right]\notag \\
&= \sqrt{\frac{H^2(t_k)}{2 k^3}} \left(\frac{S(t_k)}{S(t)}\right)^{1/2}  \exp\left[-\frac{9}{4m^2}\left(H(t_k)-\frac{2m}{3} \sqrt{c-\frac{1}{2}}\right)^2 + \frac{9}{4m^2}\left(H(t)-\frac{2 m}{3} \sqrt{c-\frac{1}{2}}\right)^2 \right]. \label{pshiggs}
\end{align}
Here we neglect the term higher than $H(t_k)^3-H(t)^3$, which are suppressed by numerical factors that are smaller than $\CO(10^{-1})$. 
The exponential factor in the power spectrum of the Higgs field has a peak
at the scale $k/a(t_k) \simeq (2/3) m\sqrt{c-1/2}$, which characterizes the power spectrum. 
At larger scales, the power spectrum decays exponentially.

From Eq.~\eqref{pshiggs}, the power spectrum of the Higgs field at $t=t_{\rm f}$ is estimated as 
\begin{equation}
{\cal P}_\sigma = |\sigma_{k}|^2 \simeq \frac{H^2(t_k)}{2k^3} \left(\frac{S(t_k)}{S(t)}\right)\exp \left[-\frac{9}{2m^2}\left(H(t_k)-\frac{2 m}{3}\sqrt{c-\frac{1}{2}}\right)^2+\frac{(2-\sqrt{3})^2c}{4}\right]. 
\end{equation}
The power spectrum decays quickly at the scale that satisfies 
\begin{equation}
\exp \left[-\frac{9}{2m^2}\left(H(t_k)-\frac{2 m}{3}\sqrt{c-\frac{1}{2}}\right)^2 \right] \ll 1. 
\end{equation}
At larger scale, any structure will not appear. 
Quantitatively,  from the discussion of Ref. \cite{Nagasawa:1991zr}, we conjecture that the largest structure will form at
the scale satisfying 
\begin{equation}
\exp \left[-\frac{9}{2m^2}\left(H(t_k)-\frac{2 m}{3}\sqrt{c-\frac{1}{2}}\right)^2 \right] \sim 10^{-2}.
\end{equation}
This corresponds to the scale $k_c$ satisfying 
\begin{equation}
H(t_{k_{\rm c}})\simeq m\left(\frac{2 }{3}\sqrt{c-\frac{1}{2}}+1\right). 
\end{equation}
In other words, 
the power spectrum decays at $k<k_{\rm c}$. 

Now we estimate the distribution and number density of monopoles. 
For this purpose, we first consider the distribution of a massless scalar field in the de Sitter background 
with Hubble parameter $H$.  
Suppose that a massless scalar field $\chi$ takes $\chi=0$ at $t=0$ uniformly. 
This leads to the Gaussian distribution of $\chi$ at $t>0$ 
and scale-invariant spectrum, $|\chi(k)|^2 = H^2/2k^3$. 
Then, the two-point probability distribution function reads, 
\begin{align}
\rho_2&[\chi({\vecs x}_1,t)=\chi_1, \chi({\vecs x}_2, t)=\chi_2]\notag \\
&=\frac{1}{2 \pi G(0,t)\sqrt{1-G^2(r,t)/G^2(0,t)}} \exp\left\{ -\frac{\chi_1^2+\chi_2^2-2 (G(r,t)/G(0,t)) \chi_1 \chi_2}{2 G(0,t)[1-G^2(r,t)/G^2(0,t)]}\right\}, 
\end{align}
where 
\begin{align}
G(0,t)&= \langle \chi^2({\vecs x},t)\rangle =\frac{H^3 t}{4 \pi^2}, \\
G(r,t)&= \langle \chi ({\vecs x}_1,t) \chi({\vecs x}_2,t)\rangle = \frac{H^3 t}{4 \pi^2} \left(1-\frac{1}{Ht}\log(Hr)\right), \quad r \equiv |{\vecs x}_1-{\vecs x}_2|>H^{-1}. 
\end{align}
Since $\rho_2$ expresses the probability at $t$ that the value of $\chi$ is $\chi_1$ at ${\vecs x}={\vecs x}_1$ and 
$\chi_2$ at ${\vecs x}={\vecs x}_2$, any correlation function $\langle F[\chi({\vecs x}_1,t),\chi({\vecs x}_2,t)]\rangle$ 
is written by 
\begin{equation}
\langle F[\chi({\vecs x}_1,t),\chi({\vecs x}_2,t)]\rangle = \int_{-\infty}^\infty d \chi_1   \int_{-\infty}^\infty d \chi_2 F[\chi_1, \chi_2] \rho_2[ \chi_1, \chi_2, t]. 
\end{equation}   
Note that a monopole exists between two separate points if all the signs of three scalar fields 
are opposite at these two points. 
Then, the probability of existence of monopoles between ${\vecs x}_1$ and ${\vecs x}_2$ is
\begin{align}
P(t) &=\left(\int_{-\infty}^0 d\chi_1 \int_0^{\infty} d\chi_2 2  \rho_2[\chi_1, \chi_2,t] \right)^3 
=\frac{1}{\pi^3} \left\{\cos^{-1}\left(\frac{G(r,t)}{G(0,t)}\right)\right\}^3\notag \\
&=\frac{1}{\pi^3} \left\{\cos^{-1}\left(1-\frac{1}{Ht}\log(Hr)\right)\right\}^3. 
\end{align}
For $Ht\gg 1$ and $Hr \simeq e$, it can be approximated as 
\begin{equation}
P(t) \simeq \frac{2^{3/2}}{\pi^3 (Ht)^{3/2}}. \label{pt1}
\end{equation}

Let us relate $P(t)$ to the distribution of monopoles. 
Define $n(V)$ as the number density of monopoles. 
Here $V\sim l^3$, where $l$ is the mean separation of monopoles. 
In other words, $V$ is the volume that a monopole occupies. 
The possible value of $V$ is $H^{-3}\lesssim V \lesssim H^{-3} e^{3Ht}$. 
$P(t)$ can be understood as the probability that {\it there is a monopole within a distance of $e/H$}. 
Thus, it can be expressed as 
\begin{equation}
P(t) \simeq e^3 H^{-3}\int_{H^{-3}}^{H^{-3} e^{3 Ht}} n(V,t) dV.  \label{pt2}
\end{equation}
Comparing Eq. \eqref{pt1} and Eq. \eqref{pt2}, we arrive at the relation,  
\begin{equation}
n(V,t) \propto V^{-\alpha}, \quad \alpha \simeq 1, 
\end{equation}
for large $Ht$. 
Therefore, we conclude that if a Higgs field has a scale-invariant power spectrum, 
the resultant number density of monopoles would be $n(V) \propto V^{-1}$. 
The possible value of $V$ is determined by the scales where the scale invariance holds.

From the discussion above, the average number density of monopoles can be estimated as 
\begin{align}
n_{\rm m}(t_{\rm f}) &\simeq \int_{H_{\rm f}^{-3}}^{(k_{\rm c}/a(t_{\rm f}))^{-3}} n(V,t_{\rm f}) dV \simeq  \frac{\int_{H_{\rm f}^{-3}}^{(k_{\rm c}/a(t_{\rm f}))^{-3}} V^{-1} dV}{\int_{H_{\rm f}^{-3}}^{(k_{\rm c}/a(t_{\rm f}))^{-3}} dV}= \frac{3 \log(H_{\rm f} a(t_{\rm f})/k_c)}{(k_{\rm c}/a(t_{\rm f}))^{-3} -H_{\rm f}^{-3}}. 
\end{align}
Note that $k/a(t_{\rm f})$ is written as
\begin{equation}
\frac{k}{a(t_{\rm f})}= \frac{a(t_k)}{a(t_{\rm f})} \frac{k}{a(t_k)} =\exp \left[-\frac{3}{2m^2}(H(t_k)^2-H(t_{\rm f})^2)\right] H(t_k). 
\end{equation}
Exponential expansion of the Universe continues after the phase transition until inflation ends at 
$H\simeq H_{\rm e}=\sqrt{1/6} m \varphi_{\rm e} /M_G \simeq m/\sqrt{3}$. 
The number of $e$-folds between the phase transition and the end of inflation 
has been estimated in Eq.~\eqref{efolaft}. 
Therefore, monopoles produced at the phase transition are diluted and 
the average number density of monopoles at the end of inflation is estimated as 
\begin{equation}
n_{\rm m} (t_{\rm e})  \simeq n_{\rm m} ( t_{\rm f})\exp\left[-\dfrac{3}{2}(c-\sqrt{2c}-1)\right]. 
\end{equation}
At larger $c$, $n_{\rm m}(t_{\rm e})$ behaves as 
\begin{equation}
n_{\rm m}(t_{\rm e}) \simeq \frac{4 m^3 c^{5/2}}{27} e^{-2c}=\frac{4 (\kappa M)^5}{27 m^2} \exp\left[-\frac{2 \kappa^2 M^2}{m^2}\right].
\end{equation}

\subsection{Constraints on the model parameter}

Let us estimate the present monopole abundance. 
After inflation, the inflaton starts damped oscillation and the Universe expands like the matter-dominated era. 
Eventually, the inflaton decays into radiation and the Universe is reheated. 
The monopole-to-entropy ratio is fixed after the reheating, and it is estimated as 
\begin{equation}
\frac{n_{\rm m}}{s} (t_R)\simeq \frac{3}{4} \left(\frac{g_*}{g_{s*}}\right)\frac{T_R}{m^2 M_G^2} n_{\rm m} (t_{\rm e}),  \label{nsratio}
\end{equation}
where $T_R$ is the reheating temperature and $g_* \simeq g_{s*} \simeq 200$ 
are the relativistic degrees of freedom. 
If there are no late-time entropy production processes, this quantity is conserved until the present time. 

The abundance of monopoles are constrained by the condition that it must not exceed 
the dark matter abundance. 
The present dark matter abundance is given by~\cite{Komatsu:2010fb}
\begin{equation}
\Omega_{\rm DM} h^2\simeq 0.11, 
\end{equation}
where $h\equiv H_0/(100 \ {\rm km \ sec}^{-1} {\rm Mpc}^{-1})\approx 0.7$ and $H_0$ is the present Hubble parameter.  
This can be rewritten in terms of the dark matter energy density-to-entropy ratio, 
\begin{equation}
\frac{\rho_{\rm DM}}{s}  \simeq 4.1 \times 10^{-10}\, \text{GeV}. \label{dmabb}
\end{equation}
On the other hand, from Eqs.~\eqref{massmp} and \eqref{nsratio}, the present monopole abundance is 
estimated as 
\begin{align}
\frac{\rho_{\rm m}}{s} 
= \frac{M_{\rm m} n_{\rm m}}{s}\simeq&  3.2 \times 10^{-8} \left(\frac{M}{10^{15}{\rm GeV}}\right) \left(\frac{0.5}{g_{\rm G}}\right)\left(\frac{g_*}{g_{*s}}\right)\left(\frac{m}{10^{13}{\rm GeV}}\right)\left(\frac{n_{\rm m}(t_{\rm e})}{m^3}\right)  T_R. 
\end{align}
Therefore, the constraint on the monopole abundance is expressed as 
\begin{align}
T_R < 
1.3 \times 10^{-2}  {\rm GeV} &\times \left(\frac{M}{10^{15}{\rm GeV}}\right)^{-1}  \left(\frac{g_{\rm G}}{0.5}\right)\left(\frac{g_{*s}}{g_*}\right)\left(\frac{m}{10^{13}{\rm GeV}}\right)^{-1} \left(\frac{n_{\rm m}(t_{\rm e})}{m^3}\right)^{-1}. 
\end{align}

Next we consider the constraint from the flux of monopoles. 
The average number density of monopoles estimated above corresponds to the flux of monopoles as
\begin{align}
F=\frac{n_{\rm m} v_{\rm m}}{4 \pi} 
\simeq & 9.1 \times 10^{-9} {\rm cm}^{-2} {\rm sr}^{-1} {\rm s}^{-1} \notag \\
&\times \left(\frac{\beta_{\rm m}}{10^{-3}}\right) \left(\frac{g_*}{g_{*s}}\right)\left(\frac{m}{10^{13}{\rm GeV}}\right)  \left(\frac{T_R}{10^{6}{\rm GeV}}\right)\left(\frac{n_{\rm m}(t_{\rm e})}{m^3}\right), 
\end{align}
where $v_{\rm m} \equiv \beta_{\rm m} c$ is the average velocity of monopoles. 
Monopoles are accelerated by the gravitational or magnetic field of our galaxy. 
Gravitational field can accelerate monopoles up to the virial velocity, $\beta_{\rm m} \sim 10^{-3}$, and 
magnetic field can accelerate them up to $\beta_{\rm m} \sim 10^{-3} (M_{\rm m}/10^{17}{\rm GeV})^{1/2}$ \cite{Parker:1970xv}. 
Here we treat it as a parameter. 
If monopoles are heavy, $M_{\rm m}\gtrsim10^{17}{\rm GeV}$, the galactic magnetic field cannot accelerate monopoles 
above the virial velocity and monopoles are 
clumped in galaxies. 
In this case, the number density and the flux of monopoles are enhanced by the factor of up to $10^5$ 
\cite{Parker:1970xv}. 
For the scale we are interested in, $\kappa M \simeq 10^{13}$\,GeV, 
the monopole mass is $M_m=4 \pi M/g_{\rm G}>10^{15-17}$\,GeV, for 
$\kappa \sim 10^{-1}-10^{-3}$. 
On the other hand, the monopole flux is severely constrained from the condition that 
the galactic magnetic field is not dissipated by the monopole from the very beginning of the galaxy formation.
This is known as the extended Parker bound~\cite{Adams:1993fj}, which reads
\begin{equation}
F\lesssim 1.2 \times 10^{-16} \left(\frac{M_{\rm m}}{10^{17}{\rm GeV}}\right) {\rm cm}^{-2} {\rm sr}^{-1} {\rm s}^{-1}.
\end{equation}
This translates into the constraint on the reheating temperature, 
\begin{equation}
T_R \lesssim 3.3 \times 10^{-9}{\rm GeV} \left(\frac{g}{0.5}\right)^{-1}\left(\frac{\beta_{\rm m}}{10^{-3}}\right)^{-1} \left(\frac{g_{*s}}{g_*}\right)\left(\frac{m}{10^{13}{\rm GeV}}\right)^{-1} \left(\frac{M}{10^{15}{\rm GeV}}\right)\left(\frac{n_{\rm m}(t_{\rm e})}{m^3}\right)^{-1} , 
\end{equation}
if monopoles are distributed uniformly in the Universe, and 
\begin{equation}
T_R \lesssim 3.3 \times 10^{-14}{\rm GeV} \left(\frac{g}{0.5}\right)^{-1}\left(\frac{\beta_{\rm m}}{10^{-3}}\right)^{-1} \left(\frac{g_{*s}}{g_*}\right)\left(\frac{m}{10^{13}{\rm GeV}}\right)^{-1} \left(\frac{M}{10^{15}{\rm GeV}}\right)\left(\frac{n_{\rm m}(t_{\rm e})}{m^3}\right)^{-1} , 
\end{equation}
if monopoles are clumped in the galaxy.
Note that $n_{\rm m}(t_{\rm e})$ behaves $n_{\rm m}(t_{\rm e}) \propto c^{5/4} \exp(-2 c) = (\kappa M/m)^{5/2}  \exp (- 2 \kappa^2 M^2/ m^2)$ at large $\kappa M/m$, 
but it has more complicated dependence on $\kappa M/m$ at the parameter region where we are interested in.

Now we turn to the possibility of direct detection of monopoles \cite{Hardtke:2007zz}. 
IceCube will be able to put the severest constraints on the monopole flux. 
It has a sensitivity to nonrelativistic magnetic monopoles thorough the catalyzed nucleon decay.  
IceCube (3-year) may have a sensitivity of monopole flux \cite{Hardtke:2007zz}\footnote{For the constraints on the flux of 
relativistic monopoles, see Ref.~\cite{Giacomelli:2011re}. Latest constraints by ANTARES are given in
Ref.~\cite{ANTARES:2011xr}. For the prospect of the constraints by IceCube, see Ref.~\cite{Helbing:2011wf}.  },   
\begin{equation}
F \sim 10^{-19} {\rm cm}^{-2} {\rm sr}^{-1} {\rm s}^{-1}
\end{equation}
for $M_{\rm m}>10^{17}{\rm GeV}$. Therefore, even if the flux of monopoles are lower than the current constraints 
described above, 
they can be detected by IceCube if
\begin{equation}
T_R \gtrsim 1.1 \times 10^{-16}{\rm GeV} \left(\frac{g}{0.5}\right)^{-1}\left(\frac{\beta_{\rm m}}{10^{-3}}\right)^{-1} \left(\frac{g_{*s}}{g_*}\right)\left(\frac{m}{10^{13}{\rm GeV}}\right)^{-1} \left(\frac{n_{\rm m}(t_{\rm e})}{m^3}\right)^{-1} . 
\end{equation}
Here we have taken into account the clumped distribution of such heavy monopoles. 
In this case, we can prove  (or disprove) the scenario, 
combined with the results from CMB polarization measurements 
to determine the inflation energy scale, as well as the reheating temperature of the Universe. 

\begin{figure}[htbp]
\begin{tabular}{c}
\includegraphics[width=100mm]{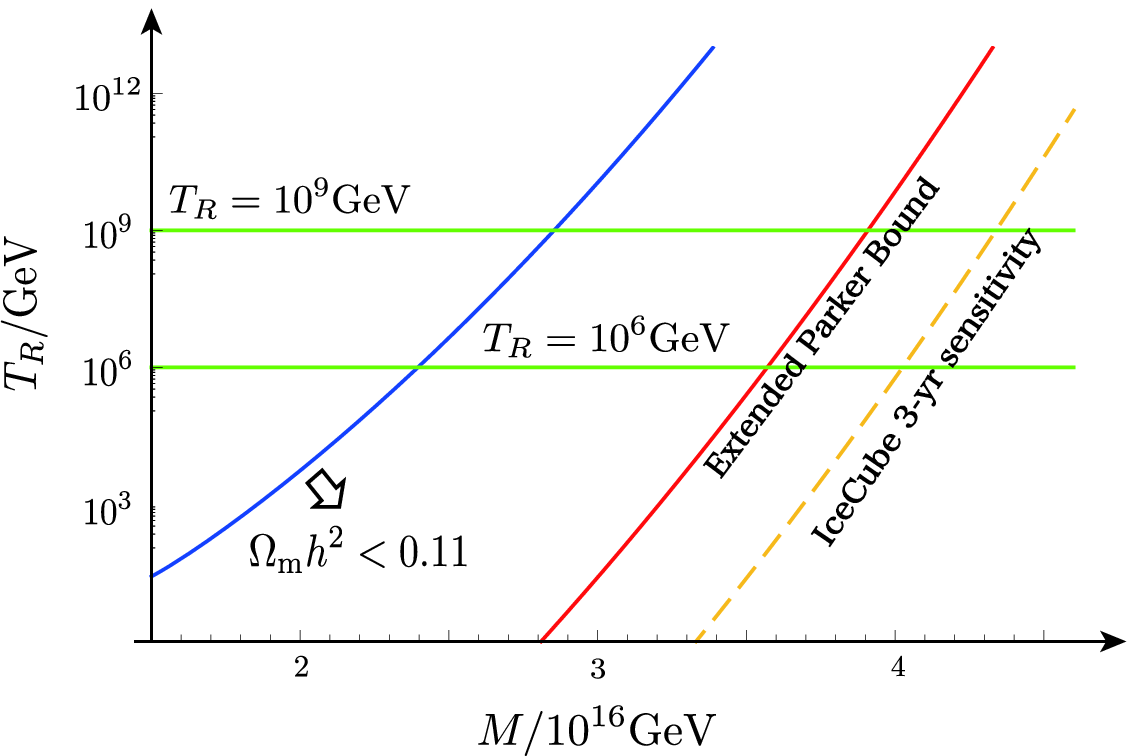} \\
 \includegraphics[width=100mm]{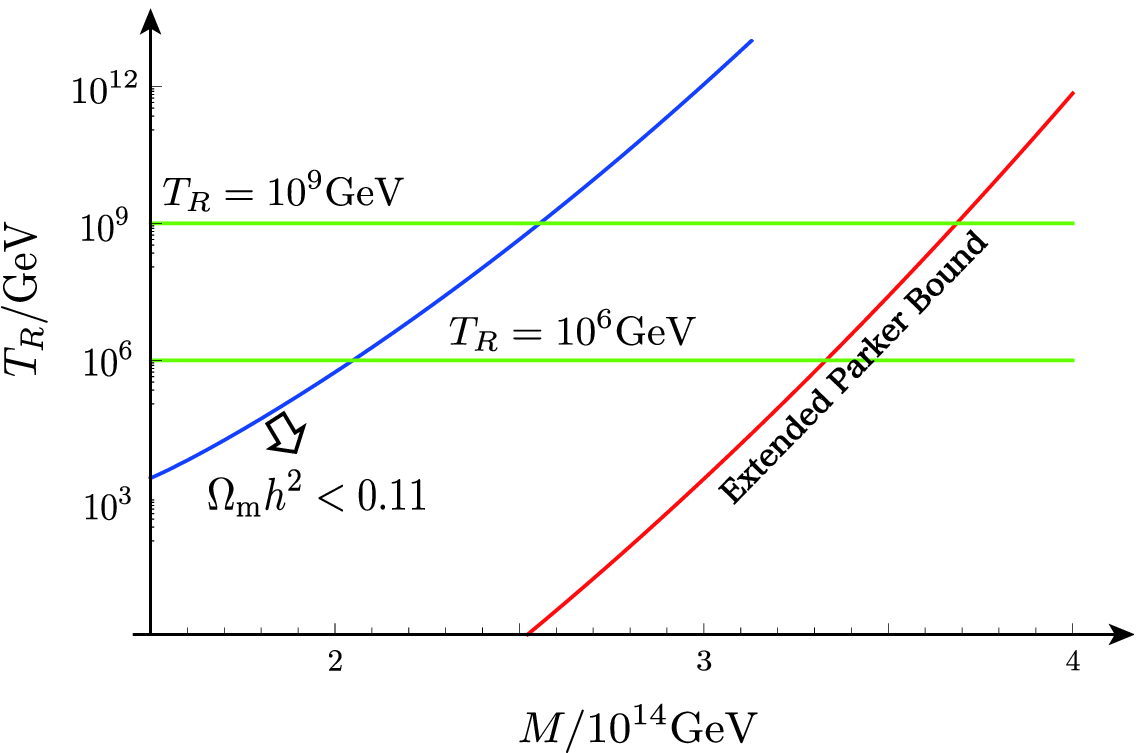}
\end{tabular}
\vspace{-0.3cm}
\caption{The allowed region of the reheating temperature (longitudinal axis) and the symmetry breaking scale $M$  (horizontal axis). Dashed (yellow) line suggests the IceCube 3-year sensitivity. Here we set $\kappa=10^{-3} (10^{-1})$ on the upper  (lower) panel  and $\beta_{\rm m}=10^{-3} (10^{-2})$.
}
\label{fig.1}
\end{figure}

\begin{figure}[htbp]
\begin{tabular}{c}
\includegraphics[width=100mm]{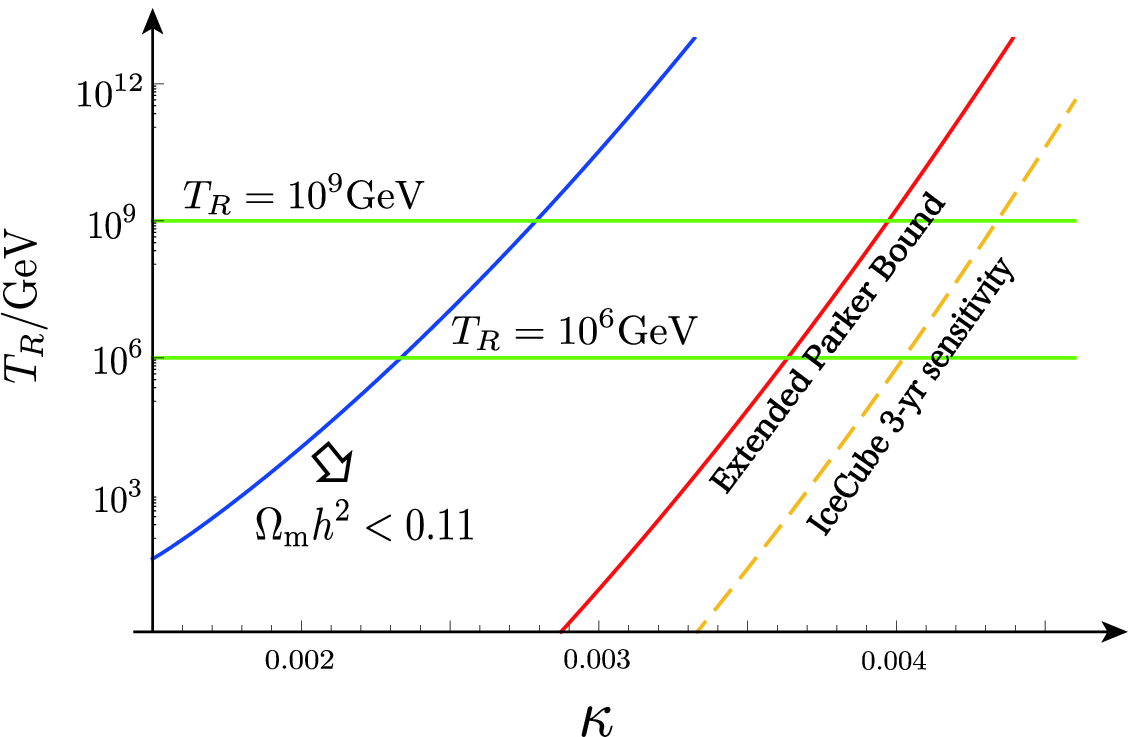} \\
 \includegraphics[width=100mm]{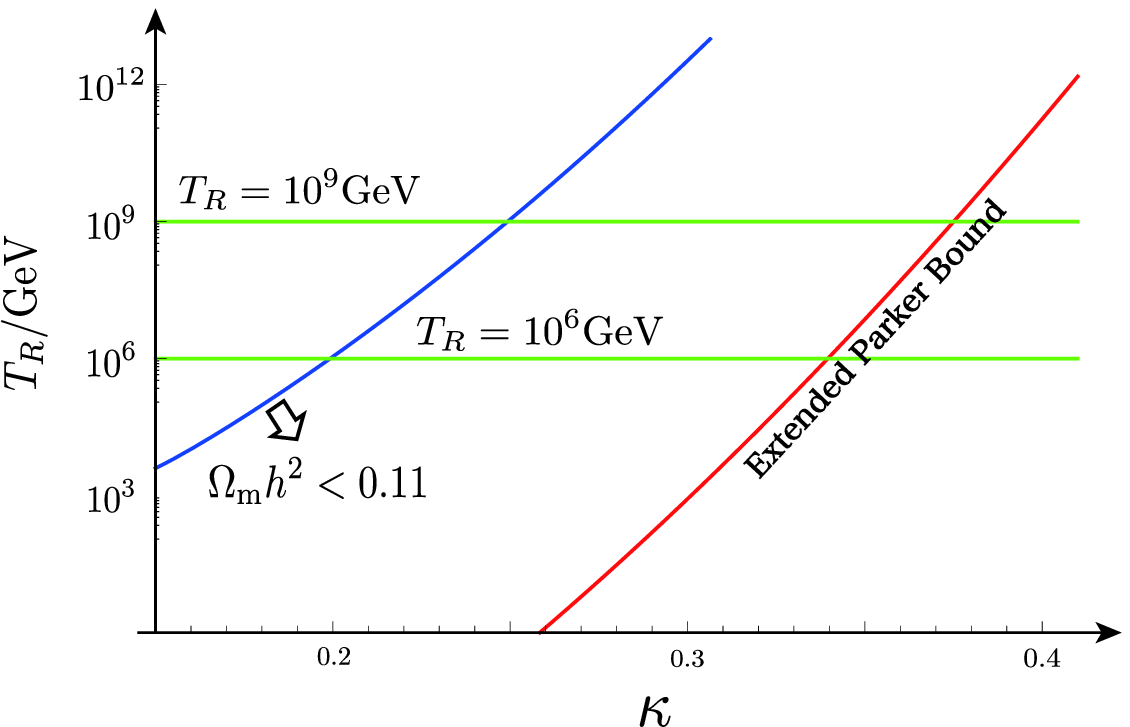}
\end{tabular}
\vspace{-0.3cm}
\caption{The allowed region of the reheating temperature (longitudinal axis) and the coupling constant $\kappa$  (horizontal axis). Dashed (yellow) line suggests the IceCube 3-year sensitivity. Here we set $M=10^{16} (10^{14}) {\rm GeV}$ on the upper  (lower) panel  and $\beta_{\rm m}=10^{-3} (10^{-2})$.}
\label{fig.2}
\end{figure}

Figures \ref{fig.1} and \ref{fig.2} show the constraints on the reheating temperature. 
In the case of $\kappa \sim 0.1$, the monopole mass is relatively small, $M_{\rm m} \sim 10^{15} {\rm GeV}$. 
In this case, the magnetic field of the galaxy accelerates monopoles up to $\beta_{\rm m} \simeq 10^{-2}$, 
which exceeds the virial velocity and they are distributed uniformly. 
IceCube may not have sensitivity to such monopoles although careful estimation is required. 
In the case of $\kappa \sim 10^{-3}$, on the other hand, the monopole mass is around 
$M_{\rm m} \sim 10^{17} {\rm GeV}$. In this case, monopoles are clumped in the galaxy. 
We can see that these conditions are stronger than the constraint from the gravitino problem, 
$T_R<10^{6-9}{\rm GeV}$~\cite{Kawasaki:2008qe}, 
at $M \simeq (2-4) \times 10^{13}{\rm GeV} \kappa^{-1} $. 
We can also see that there are parameter regions where monopole detection 
can be expected by IceCube while these constraints are avoided. 

Let us summarize the results of this section for the model given by (\ref{Kahler}) and (\ref{super}). 
i)  The case of $M < 10^{13} \kappa ^{-1}$ GeV is incompatible with the standard cosmological scenario 
because monopoles would overclose the Universe. 
Note that it is possible to dilute monopoles by some late-time entropy production processes, 
which can be verified by future gravitational wave
experiments~\cite{Nakayama:2008ip} such as DECIGO~\cite{DECIGO} or BBO~\cite{BBO}. 
ii) The case of $M \sim 10^{13}\kappa ^{-1}$ GeV is the most interesting and the flux of monopoles 
may be within the reach of IceCube~\cite{Hardtke:2007zz}, 
depending on the choice of $\kappa$ and $T_R$.  
iii) In the case of $M >10^{13}\kappa ^{-1}$ GeV, monopoles are diluted during inflation sufficiently
and there are no significant effects on cosmology. 
As a result, if the PLANCK satellite~\cite{PLANCK_HP,PLANCK_coll} or other B-mode measurements 
should favor chaotic inflation, the symmetry breaking scale would be constrained.

\subsection{Realistic Models \label{sec:5}}

Thus far, we have studied general features of phase transitions during F-term (chaotic) inflation. 
Then, a natural question is whether this mechanism can be embedded in specific models of SUSY GUTs. 
In this section, we comment on this issue using some realistic GUT models. 
We also note that in some symmetry breaking patterns, not only monopoles but also 
domain walls could be produced and the constraint would become severer than our estimate in the previous section. 

One of the simplest candidates of SUSY GUT is an $SU(5)$ model~\cite{su5}. 
In this model, we can realize the symmetry breaking $SU(5) \rightarrow SU(3)_c\times SU(2)_L \times U(1)_Y$ 
by introducing an adjoint Higgs multiplet $\Phi$ and assume a superpotential, such as 
\begin{equation}
W=S (\mu^2- \alpha {\rm Tr} \Phi^2) -\beta {\rm Tr} \Phi^3. 
\end{equation}
However, this model turns out to have separated minima with different gauge symmetries, namely, 
$SU(4) \times U(1)$ and $SU(3)\times SU(2) \times U(1)$.
In our scenario, Higgs field can fall into both of the minima. 
These minima are topologically disconnected and hence domain walls should be 
formed as well as monopoles. 
Thus we do not have a consistent cosmological evolution scenario in this case, 
unless the symmetry breaking occurs well before the comoving Hubble scale today 
left the Hubble radius during inflation. Thus we cannot hope to detect magnetic monopoles. 

On the other hand, the Pati-Salam model~\cite{Pati:1974yy}, $SU(4)_C \times SU(2)_L\times SU(2)_R$, which is broken by two 
Higgs multiplets, $H=(4,1,2)$ and ${\bar H}=({\bar 4},1,2)$, is applicable in our scenario. 
The superpotential is given by 
\begin{equation}
W=\kappa S (H {\bar H}-\mu^2).  
\end{equation}
This model has a $SU(3)_c \times SU(2)_L\times U(1)_Y$ vacuum at $|H|=|{\bar H}|=\mu$. 
For these fields, there are only three degree of freedom in the D-flat direction. 
Therefore, the pattern of symmetry breaking is unique. 
During the course of this phase transition, monopole production is inevitable and hence 
our scenario discussed in the previous section can be applied. 
It would be interesting to seek for other models with the same symmetry breaking property and investigate 
their phenomenology. This is left for future study.

\section{Conclusion and Discussion}

In this paper, we have studied phase transition driven by Hubble-induced mass in supersymmetric F-term inflation 
models\footnote{Formation of cosmic strings in the same mechanism will be discussed elsewhere~\cite{miyamoto}. }.  
We have found that in supersymmetric F-term inflation models the breaking scale of 
GUT and other intermediate symmetry group, which is associated with the production of monopoles and domain walls,
are constrained to be larger than the Hubble scale during inflation,
because the Hubble-induced mass inevitably arises in 
supersymmetric F-term inflation models~\cite{Dine:1995uk}.
If future CMB observations such as PLANCK~\cite{PLANCK_HP,PLANCK_coll}, 
QUIET+PolarBeaR \cite{Hazumi:2008zz} or LiteBIRD \cite{Hazumi:2008zz,LiteBird} 
will detect the $B$-mode polarization and determine the inflation energy scale,
we can directly constrain the symmetry breaking scale. 
As a concrete example, we have focused on the chaotic inflation model 
because its energy scale is rather close to the GUT scale
and we studied the monopole production in detail.
We have obtained constraints on the symmetry breaking scale in order for the monopole abundance
produced during the course of inflation not to contradict with observational bounds.
If the symmetry breaking takes place close to the end of chaotic inflation, 
future experiments such as IceCube~\cite{Hardtke:2007zz} may be able to discover the signatures of monopoles.

We have opened a new window to constrain GUT or other unified theories by 
considering the phase transition during inflation. 
This study relates the CMB observations to the experiments dedicated for direct detection of monopoles.
As a theoretical issue, it may be interesting to investigate phenomenological aspects of concrete 
GUT models such as the Pati-Salam model~\cite{Pati:1974yy} in the light of our suggestions.

\section*{Acknowledgments}

K.K. would like to thank W. Buchm\"uller for the useful comments.
This work was partially supported by  JSPS Grant-in-Aid for Scientific Research
Nos.\ 22244030 (K.N.) and 23340058 (J.Y.), and Grant-in-Aid for
Scientific Research on Innovative Areas No.\ 21111006 (K.N. and J.Y.).


\end{document}